\documentclass{ws-mpla}

\usepackage{subcaption}
\usepackage[super]{cite}
\usepackage{xcolor}
\usepackage[verbose,hypertexnames=false]{hyperref}
\usepackage{lipsum} 

\usepackage{color}
\definecolor{blue}{rgb}{0.0, 0.0, 1.0}
\definecolor{red}{rgb}{1.0, 0.0, 0.0}
\definecolor{royalblue}{rgb}{0.0, 0.14, 0.4}
\hypersetup{colorlinks=true,linkcolor=red, citecolor=blue, urlcolor=blue}
\usepackage{graphicx}
\usepackage{dcolumn}
\usepackage{bm}
\usepackage{amsmath,amssymb}
\usepackage{float}
\usepackage{multirow}
\usepackage{slashed}
\usepackage{xcolor}
\usepackage{braket}
\usepackage{physics}
\usepackage{multirow}
\usepackage{gensymb}
\def\orcid#1{\kern .08em\href{https://orcid.org/#1}{\includegraphics[keepaspectratio,width=0.7em]{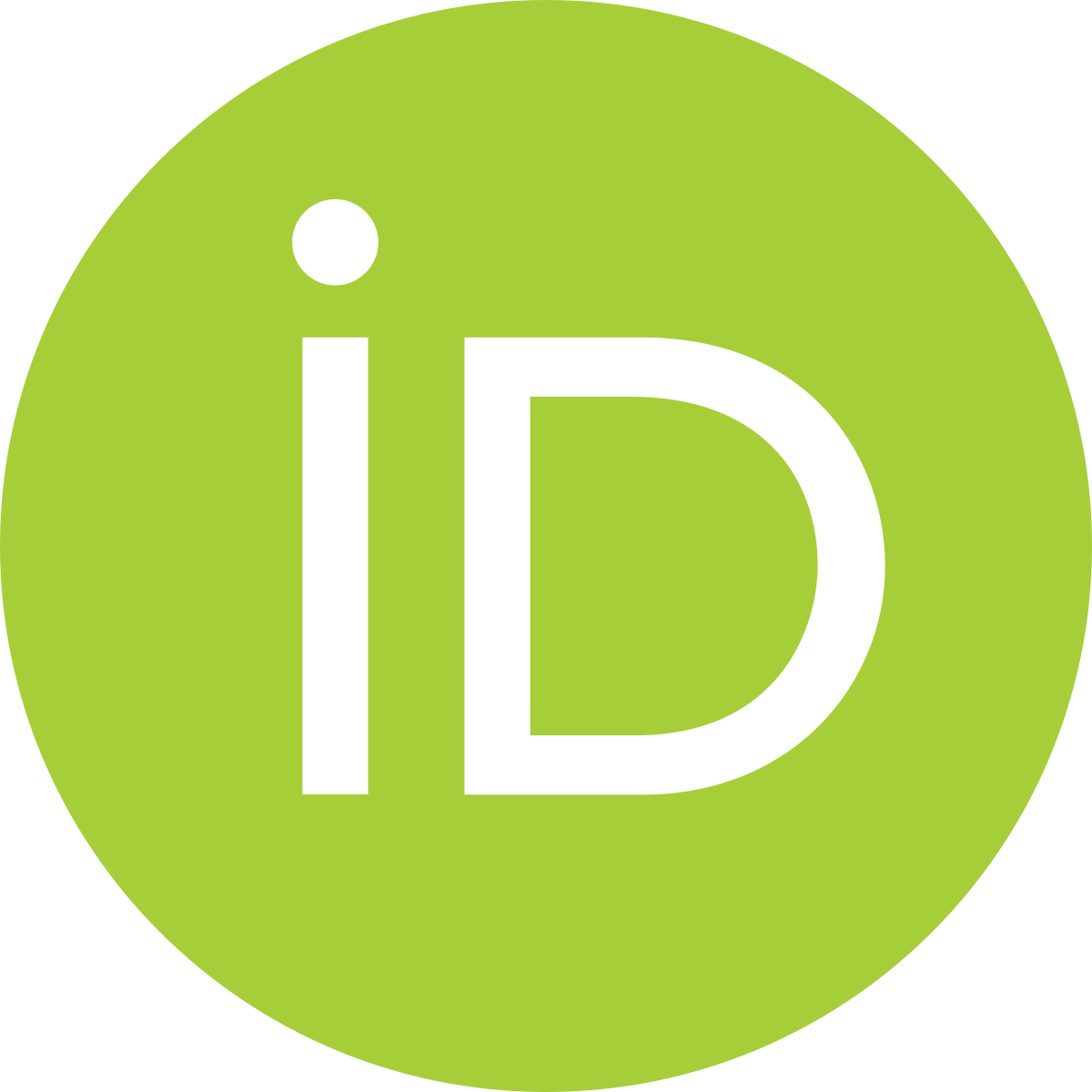}}}

\begin{document}


\catchline{}{}{}{}{}

\title{Electromagnetic structure of $B_c$ and heavy quarkonia\\ in the light-front quark model}


\author{Rayn Rasyid Harjapradipta\orcid{0009-0000-2364-9601}}
\address{Departemen Fisika, FMIPA, Universitas Indonesia, Depok 16424, Indonesia\\
rayn.rasyid@ui.ac.id}
\author{Muhammad Ridwan\orcid{0000-0002-2949-5866}}
\address{Centre for Mathematical Sciences, University of Plymouth, Plymouth PL4 8AA, UK \\
Departemen Fisika, FMIPA, Universitas Indonesia, Depok 16424, Indonesia\\
muhammad.ridwan@plymouth.ac.uk}
\author{Ahmad Jafar Arifi\orcid{0000-0002-9530-8993}}
\address{Advanced Science Research Center, JAEA, Ibaraki 319-1195, Japan\\
aj.arifi01@gmail.com}
\author{Terry Mart\orcid{0000-0003-4628-2245}}
\address{Departemen Fisika, FMIPA, Universitas Indonesia, Depok 16424, Indonesia\\
terry.mart@sci.ui.ac.id}

\maketitle

\begin{history}
\received{(Day Month Year)}
\revised{(Day Month Year)}
\accepted{(Day Month Year)}
\published{(Day Month Year)}
\end{history}

\begin{abstract}
We investigate the electromagnetic structure of heavy quarkonia and the $B_c$ meson within the light-front quark model (LFQM) to better understand the internal spatial charge distributions and QCD dynamics of heavy mesons. 
The light-front wave functions (LFWFs) are obtained using a variational approach with a few set of harmonic oscillator basis functions, providing a flexible yet tractable description of the bound-state dynamics. 
Using these LFWFs, we compute the electromagnetic form factors and compare our results with available lattice QCD data and other model calculations. 
Our results are roughly consistent with previous model predictions, showing that the electromagnetic radii of the $2S$ and $3S$ states are approximately 1.5 times and 1.9 times larger than those of their corresponding $1S$ states, reflecting the expected growth of spatial size in radial excitations.

\end{abstract}


PACS Nos.:

\section{Introduction}

Understanding the structure of hadrons, governed by quantum chromodynamics (QCD), remains a central challenge due to its inherently nonperturbative nature~\cite{Gross:2022hyw}. 
Since the discovery of charmonia in 1974~\cite{E598:1974sol}, heavy quarkonia have served as an ideal system for testing QCD models because of their relatively simple heavy quark--antiquark structure~\cite{Brambilla:2010cs,Eichten:2007qx,Voloshin:2007dx}. 
In addition, the spectrum, decay rates, and transition properties of heavy quarkonia provide sensitive probes of the strong interaction, including spin-dependent forces, relativistic corrections, and confinement dynamics. 

Among the various properties that probe the internal dynamics of heavy quarkonia and the $B_c$ meson, electromagnetic form factors (EMFFs) play a crucial role. 
They provide direct information on the spatial distribution of charge and current within the meson, and thus serve as an important link between the experiment, lattice QCD~\cite{Lewis:2012bh, Li:2020gau, Dudek:2006ej, Delaney:2023fsc}, and phenomenological models including the Basis Light-Front Quantization (BLFQ) and the Bethe-Saltpeter Equation (BSE)~\cite{Arifi:2022pal, Li:2015zda,Hernandez-Pinto:2023yin, Adhikari:2018umb, Serafin:2020egn,Hwang:2001th, Zhang:2024nxl, Maris:2000sk,Wang:2022mrh}. 
While EMFFs of light mesons can be accessed experimentally~\cite{JeffersonLab:2008jve,Amendolia:1984nz}, 
for heavy quarkonia and the $B_c$ meson, direct experimental measurements are currently unavailable due to their short lifetimes and production challenges. 
In this context, lattice QCD calculations offer a valuable benchmark, allowing us to test and validate phenomenological models.


Among phenomenological models, the light-front quark model (LFQM)~\cite{Choi:2007se,Choi:2009ai} has been extensively used to study meson properties because light-front dynamics (LFD)~\cite{Dirac:1949cp,Brodsky:1997de} naturally incorporates relativistic effects due to its unique rational energy-momentum dispersion relation.  In comparison with other models, the BLFQ solves the light-front Hamiltonian nonperturbatively in a confined basis. Meanwhile for the BSE framework, the model provides a covariant bound-state equation from which the light-front wave function is obtained by projecting the BSE amplitude onto the light front. In contrast to these methods, the LFQM model uses fixed trial light-front wavefunctions, while the lattice QCD uses the Feynman path integral in the action and solve it numerically. In other words, it can provide a clear partonic picture, and allows the study of the three-dimensional structure of hadrons.

Previous LFQM studies have extensively investigated various meson properties, including mass spectra, decay processes, form factors, and partonic structures\cite{Choi:1997iq,Choi:2015ywa,Acharyya:2024tql}. 
Moreover, self-consistent analyses using different current components and polarization states have also been carried out recently, and these studies have been extended to various processes. 
In particular, the $M \to M_0$ replacements based on the Bakamjian-Thomas (BT) construction allow for self-consistent results. 
For instance, the transition form factor in the $M1$ radiative transition with different currents and polarizations has been examined~\cite{Ridwan:2024ngc}.

The light-front wave functions (LFWFs) have been constructed variationally using harmonic oscillator basis states~\cite{Arifi:2022pal} for higher radial excitations~\cite{Ridwan:2024ngc}. 
Building on our previous work~\cite{Ridwan:2024ngc}, we aim to investigate the electromagnetic form factors (EMFFs) of charmonia $(c\bar{c})$, $B_c$ mesons $(b\bar{c})$, and bottomonia $(b\bar{b})$, including states up to the second radial excitation, within the LFQM. 
The resulting predictions are compared with available lattice QCD data to test the model and provide further insight into the electromagnetic structure of heavy-flavor hadrons.

This paper is organized as follows. In section-~\ref{sec:LFQM}, we present the formulation of the LFQM used in this study. 
The formulation for computing the EMFFs is given in   section-~\ref{sec:EMFF}.
In section-~\ref{sec:results}, we discuss our numerical results for the EMFFs and charge radii, comparing them with available lattice QCD data.
Finally, we summarize our findings in section-~\ref{sec:conclusion}.

\section{Light-Front Quark Model}
\label{sec:LFQM}

In this section, we present the basic idea of the LFQM for describing the properties and structure of mesons~\cite{Choi:1997iq,Choi:2015ywa,Ridwan:2024ngc,Arifi:2022pal}. 
The radial wave function is treated as a trial wave function using the harmonic oscillator (HO) basis function for the variational analysis of the QCD-motivated effective Hamiltonian, which saturates the Fock state expansion with the constituent quark and antiquark. 
The meson system at rest is described as a bound system of dressed valence quark and antiquark satisfying the eigenvalue equation of the QCD-motivated effective Hamiltonian, $H_{q\bar{q}} \ket{\Psi_{q\bar{q}}} = M_{q\bar{q}} \ket{\Psi_{q\bar{q}}},$
where $M_{q\bar{q}}$ and $\Psi_{q\bar{q}}$ are the mass eigenvalue and eigenfunction of the $q\bar{q}$ meson state, respectively. 
We take the Hamiltonian $H_{q\bar{q}} = H_0 + V_{q\bar{q}}$, where
$H_0$ is the kinetic energy part of the quark and antiquark in the relativistic form. 
The effective inter-quark potential $V_{q\bar{q}}$ includes the confinement, Coulomb, and spin-spin potential~\cite{Ridwan:2024ngc}.

The LFWF is represented by the Lorentz invariant internal variables 
\begin{eqnarray}
x_i = p^+_i /P^+, \qquad \bm{k}_{\perp i} = \bm{p}_{\perp i} - x_i \bm{P}_{\perp},   
\end{eqnarray}
and helicity $\lambda_i$, where 
$P^\mu = (P^+,P^-,\bm{P}_\perp)$ is the four-momentum of the meson, and $(p^\mu_q, p^\mu_{\bar{q}})$ are the four-momentum of the quark and antiquark, respectively. 
This leads to the constraints $x_q + x_{\bar{q}} = 1$ and 
$\bm{k}_{\perp q} + \bm{k}_{\perp \bar{q}} = 0$.
Here we assign $x \equiv x_q$ with $\bm{k}_\perp \equiv \bm{k}_{\perp q}$.
The LFWF of the $nS$ state pseudoscalar and vector mesons in the momentum space is then given by 
\begin{eqnarray}\label{eq:LFWF}
		\Psi^{Jh}_{nS}(x, \bm{k}_{\bot},\lambda_i) = \Phi_{nS}(x, \bm{k}_\bot)
		\  \mathcal{R}^{Jh}_{\lambda_q\lambda_{\bar{q}}}(x, \bm{k}_\bot),
\end{eqnarray}
where $\Phi_{nS}$ is the radial wave function and $\mathcal{R}^{Jh}_{\lambda_q\lambda_{\bar{q}}}$ 
is the spin-orbit wave function that is obtained by the interaction-independent Melosh transformation~\cite{Melosh:1974cu} from the ordinary
spin-orbit wave function assigned by the quantum number $J^{PC}$. In this work, we only consider the pseudoscalar mesons with $\mathcal{R}^{Jh}_{\lambda_q\lambda_{\bar{q}}}$ given by
\begin{eqnarray}
	\mathcal{R}^{Jh}_{\lambda_q\lambda_{\bar{q}}} &=& \frac{1}{\sqrt{2} \tilde{M}_0} 
	\bar{u}_{\lambda_q}^{}(p_q) \gamma_5 v_{\lambda_{\bar{q}}}^{}(p_{\bar{q}}), 
\end{eqnarray}
where $\tilde{M}_0 \equiv \sqrt{M_0^2 - (m_q -m_{\bar{q}})^2}$ and
\begin{eqnarray}
	M_0^2 = \frac{\bm{k}_{\bot}^2 + m_q^2}{x}  + \frac{\bm{k}_{\bot}^2 + m_{\bar{q}}^2}{1-x}.
\end{eqnarray}
Note that $\mathcal{R}^{Jh}_{\lambda_q\lambda_{\bar{q}}}$ follows the normalization $\sum_{\lambda_q,\lambda_{\bar{q}}} \braket{\mathcal{R}^{Jh}_{\lambda_q\lambda_{\bar{q}}} }{\mathcal{R}^{J^\prime h^\prime}_{\lambda_q\lambda_{\bar{q}}}} = \delta_{JJ^\prime}\delta_{h h^\prime}.$
In the radial wave function of Eq.~\eqref{eq:LFWF}, we expand the basis in the HO basis functions up to the $3S$ state,~\cite{Ridwan:2024ngc} where the HO basis functions up to the $3S$ state are expressed as
\begin{eqnarray}
	\phi_{1S}^\mathrm{HO} (\bm{k}) &=& \frac{1}{ \pi^{3/4}\beta^{3/2}} e^{-k^2/ 2\beta^2},\\
	\phi_{2S}^\mathrm{HO} (\bm{k}) &=& \frac{(2k^2 -3\beta^2)}{\sqrt{6} \pi^{3/4}\beta^{7/2}} e^{-k^2/ 2\beta^2},\\
 	\phi_{3S}^\mathrm{HO} (\bm{k}) &=& \frac{(15\beta^4 -20\beta^2k^2 + 4k^4)}{2\sqrt{30} \pi^{3/4}\beta^{11/2}}  e^{-k^2/ 2\beta^2}, \quad \quad 
\end{eqnarray}
with $k=|\bm{k}|$ and $k^2 = \bm{k}_\perp^2 + k_z^2$. The $\beta$ parameter is common for the $\phi_{nS}^\mathrm{HO}$ as it controls the width of the wave function making the basis functions orthogonal to each other.
Then, we perform a variable transformation $k_z \to x$ given by 
\begin{eqnarray}\label{eq:k_z}
    k_z = \left( x - \frac{1}{2} \right) M_0 + \frac{(m^2_{\bar{q}} -m^2_q)}{2M_0}.
\end{eqnarray} 
Thus, the complete radial wave function can be written as
\begin{eqnarray}
    \Phi_{nS}(x,\bm{k}_\bot) =  \sqrt{2(2\pi)^3} \sqrt{\frac{\partial k_z}{\partial x}}   \phi_{nS}(\bm{k}),
\end{eqnarray}
where the Jacobian factor 
\begin{equation}
\frac{\partial k_z}{\partial x} = \frac{M_0}{4x(1-x)} \left[ 1 - \frac{ (m_q^2 - m_{\bar{q}}^2)^2}{M_0^4} \right],
\end{equation}
should be included due to the variable transformation and is important to maintain the rotational symmetry~\cite{Arifi:2022qnd}.
We note that the LFWF follows the orthonormal condition as
\begin{eqnarray}
 \int \frac{\dd x\ \dd^2 \bm{k}_\bot}{2(2\pi)^3} \Psi_{nS}^{Jh\dagger} \Psi_{n^\prime S}^{J^\prime h^\prime}  = \delta_{JJ^\prime}\delta_{h h^\prime}\delta_{nn^\prime}.
\end{eqnarray}

\section{Electromagnetic Form Factor}
\label{sec:EMFF}

The electromagnetic form factor (EMFF) is defined through the matrix element of the quark current\cite{Choi:1997iq, Hwang:2001th}
\begin{eqnarray}\label{eq:formfactor}
\mel{P'}{\bar{q}\gamma^\mu q}{P} = \mathcal{P}^\mu F_\mathrm{P}(q^2)
\end{eqnarray}
where $\mathcal{P}^\mu=(P+P')^\mu$.
To compute the pseudoscalar form factor above, we use the Drell-Yan-West ($q^+=0$) frame
with $\bm{P}_\perp =0$, where $q^2=-\bm{q}^2_\perp\equiv -Q^2$. In this frame, we have
\begin{align}
    P& = \left(P^+, \frac{M^2}{P^+}, 0_\perp \right), \,\, P' =\left(P^+,  \frac{M'^2 + \bm{q}^2_\perp}{P^+}, -\bm{q}_\perp \right),\\
    q &= \left (0, \frac{M^2-M'^2 - \bm{q}^2_\perp}{P^+}, \bm{q}_\perp \right).
\end{align}


For $P(q_1{\bar q})\to P' (q_2{\bar q})$ transition with the momentum transfer $q = P-P'$, 
the relevant on-mass shell quark momentum variables in the $q^+=0$ frame are given by
\begin{align}
p^+_1 =x P^+, \,\,\,\,  p^+_{\bar q}= (1-x) P^+, \,\,\,\, \bm{p}_{1\perp} = x \bm{P}_\perp - \bm{k}_\perp, \,\,\,\, \bm{p}_{{\bar q}\perp} = (1-x) \bm{P}_\perp + \bm{k}_\perp, 
\end{align}
and 
\begin{align}
p^+_2 =x P^+, \,\,\,\, p'^+_{\bar q}= (1-x) P^+, \,\,\,\, \bm{p}_{2\perp} = x \bm{P}'_\perp - \bm{k}'_\perp, \,\,\,\, \bm{p}'_{{\bar q}\perp} = (1-x) \bm{P}'_\perp + \bm{k}'_\perp.
\end{align}
Since the spectator quark (${\bar q}$) requires that $p^+_{\bar q}=p'^+_{\bar q}$ and $\bm{p}_{{\bar q}\perp}=\bm{p}'_{{\bar q}\perp}$,
one obtains $\bm{k}'_\perp = \bm{k}_\perp + (1-x) \bm{q}_\perp$. 

The electromagnetic matrix element in the one-loop contribution $\mathcal{J}^\mu_\mathrm{em}= \mel{P'}{\bar{q}\gamma^\mu q}{P},$
within the LFQM framework based on the non-interacting $q{\bar q}$ representation consistent with the BT construction
is then obtained by the convolution of the initial and final state LF wave functions as follows:
\begin{eqnarray}
\mathcal{J}^\mu_\mathrm{em} &=&  \int^1_0 \dd p^+_1 \int \frac{\dd^2 \bm{k}_\bot}{16\pi^3}\  \Phi'(x,\bm{k}^\prime_\perp)  \Phi(x,\bm{k}_\perp)  
\sum_{\lambda_1,\lambda_2,\bar{\lambda}} h^\mu_{\lambda_1{\bar{\lambda}}\to\lambda_2{\bar{\lambda}}},
\end{eqnarray}
where
\begin{eqnarray}
\sum_{\lambda_1,\lambda_2,\bar{\lambda}} h^\mu_{\lambda_1{\bar{\lambda}}\to\lambda_2{\bar{\lambda}}} \equiv 
\sum_{{\bar{\lambda}}} (h^{\mu}_{\uparrow{\bar{\lambda}}\to\uparrow{\bar{\lambda}}}+ h^{\mu}_{\downarrow{\bar{\lambda}}\to\downarrow{\bar{\lambda}}}) + 
\sum_{{\bar{\lambda}}} (h^{\mu}_{\uparrow{\bar{\lambda}}\to\downarrow{\bar{\lambda}}}+ h^{\mu}_{\downarrow{\bar{\lambda}}\to\uparrow{\bar{\lambda}}}),
\end{eqnarray}
are the helicity non-flip and the helicity flip contributions, respectively.
Each contribution can be computed by 
\begin{eqnarray}
 h^\mu_{\lambda_1{\bar{\lambda}}\to\lambda_2{\bar{\lambda}}} \equiv \mathcal{R}^\dagger_{\lambda_2{\bar{\lambda}}}
\left[\frac{\bar{u}_{\lambda_2}(p_2)}{\sqrt{p^+_2}} \gamma^\mu \frac{u_{\lambda_1}(p_1)}{\sqrt{p^+_1}}\right]\mathcal{R}_{\lambda_1{\bar \lambda}}.
\end{eqnarray}

Applying the same BT construction to the Lorentz factor $\mathcal{P}^{\mu}$ in Eq.~(\ref{eq:formfactor}), we obtain the form factor $F^{(\mu)}_\mathrm{P}$ for any component of the current as
\begin{eqnarray}
F^{(\mu)}_\mathrm{P} (Q^2) =  \int^1_0 \dd p^+_1 \int \frac{\dd^2 \bm{k}_\bot}{16\pi^3}\  \Phi'(x,\bm{k}^\prime_\perp)  \Phi(x,\bm{k}_\perp) \sum_{\lambda_1,\lambda_2,\bar{\lambda}} \frac{ h^\mu_{\lambda_1{\bar{\lambda}}\to\lambda_2{\bar{\lambda}}}}{\mathcal{P^{\mu}}}.
\end{eqnarray}
It is important to note that all meson mass terms, denoted as $M^{(\prime)}$, appearing in $\mathcal{P}^{\mu}$ are substituted with $M^{(\prime)}_0$, 
where $M_0^\prime = M_0(\bm{k}_\perp\rightarrow\bm{k}^{\prime}_\perp)$ represents the invariant mass of the final state.

The one-loop contribution in the form factor for each current component $\mathcal{J}^\mu_\mathrm{em}$ is given by
\begin{eqnarray} \label{eq:operator}
F^{(\mu)}_\mathrm{P}(Q^2) = \int^1_0 \dd x \int\frac{\dd^2\bm{k}_\perp}{16\pi^3}
\frac{\Phi(x,\bm{k}_\perp)\Phi'(x,\bm{k}^{\prime}_\perp)}
{\sqrt{\mathcal{A}^2+ \bm{k}^2_\perp} \sqrt{\mathcal{A}^2+ \bm{k}^{\prime 2}_\perp}} \mathcal{O}^{(\mu)}_\mathrm{em},
\end{eqnarray}
where $\mathcal{O}^{(\mu)}_\mathrm{em}$ is defined as
\begin{eqnarray}
\mathcal{O}^{(\mu)}_\mathrm{em} &=&
\frac{\mathcal{N} P^+}{\mathcal{P^{\mu}}} 
\sum_{\lambda_1,\lambda_2,\bar{\lambda}} h^\mu_{\lambda_1{\bar{\lambda}}\to\lambda_2{\bar{\lambda}}}, 
\end{eqnarray}
and $P^+$ comes from the transformation of $\dd p^+_1= P^+ \dd x$ in Eq.~\eqref{eq:formfactor}.
This process involves isolating the common denominator factor,
$\mathcal{N}\equiv\sqrt{\mathcal{A}^2 + \bm{k}^2_\perp}\sqrt{\mathcal{A}^2 + \bm{k}'^2_\perp}$, and incorporating it into the wave functions. 
In the plus component $(\mu=+)$, the helicity flip contributions are zero, and only the helicity non-flip elements contribute. Thus, we obtain the operator for the plus component
\begin{eqnarray}
\mathcal{O}^{(+)}_\mathrm{em} &=& 2 (\mathcal{A}^2 + \bm{k}_\perp \cdot \bm{k}'_\perp).
\end{eqnarray}
There has been some efforts to verify the consistency using different current components~\cite{Choi:2024ptc}.
For $B_c$ meson, we should add the contribution from the photon coupled to the quark and antiquark to the form factor, while for $\eta_c$ and $\eta_b$ the only contribution from quark is considered otherwise they vanish.


The charge radius of a meson is obtained from the slope of its EMFF at vanishing momentum transfer. 
Thus, the charge radius can be extracted from
\begin{equation} \label{eq.charge_radii}
    \langle r^2 \rangle = -6 \left. \frac{dF(Q^2)}{dQ^2} \right|_{Q^2=0}.
\end{equation}
In this work, we consider the root mean square (rms) charge radius $\sqrt{\langle r^2 \rangle}$ for comparing to other models.

\section{Results and Discussion}
\label{sec:results}

In this section, we present and analyze our numerical results for the EMFF of charmonia, $B_c$ mesons, and bottomonia. 
In this work, we use model parameters from our previous work~\cite{Ridwan:2024ngc} that were obtained by fitting the screened Cornell potential to reproduce the meson spectra and decay constants. 
Then, $\beta$ parameter was found by matching the LFWFs to the potential model’s bound-state solutions. 
The parameter models are shown in Table~\ref{tab:parameter}.




\begin{table}[t!]
\centering
\tbl{Model parameters we used in this work adopting from our previous work~\cite{Ridwan:2024ngc}. The quark masses and $\beta$ parameters are given in units of GeV.\label{tab:parameter}}
{\tabcolsep10pt\begin{tabular}{@{}cccccccc@{}}
\toprule
$m_b$ & $m_c$ & $\beta_{c\bar{c}}$ & $\beta_{b\bar{c}}$& $\beta_{b\bar{b}}$   & $\theta_{12}$ & $\theta_{13}$ = $\theta_{23}$\\
\colrule
4.97(4) & 1.61(4) & 0.542(8)  & 0.702(9) & 1.060(11) & 12.12(30)\degree & 8.44(14)\degree \\

\botrule
\end{tabular}}
\end{table}

\subsection{Charmonia}

\begin{figure}[t!]
\centering
\begin{subfigure}{0.48\textwidth}
    \centering
    \includegraphics[width=\linewidth]{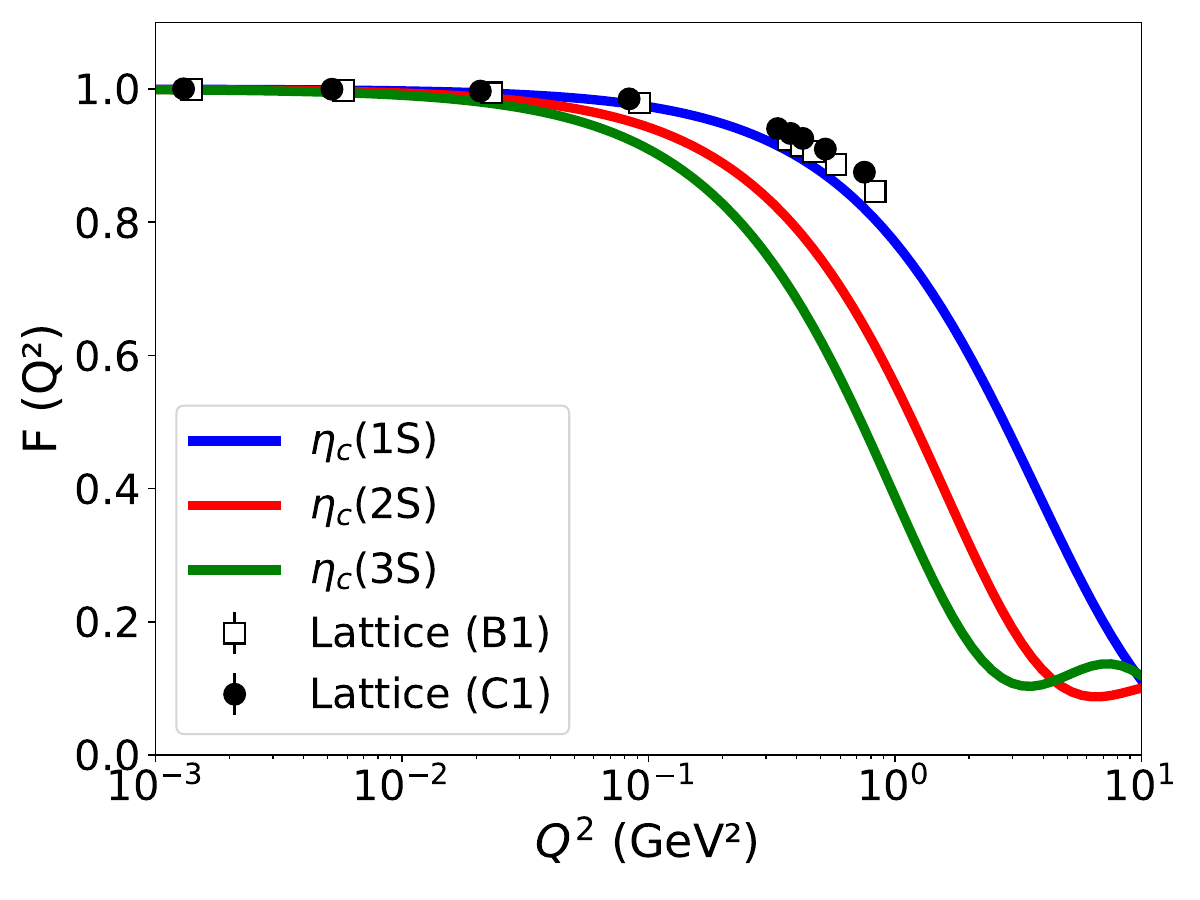}
    \subcaption{}
    \label{fig:cc}
\end{subfigure}%
\hfill
\begin{subfigure}{0.48\textwidth}
    \centering
    \includegraphics[width=\linewidth]{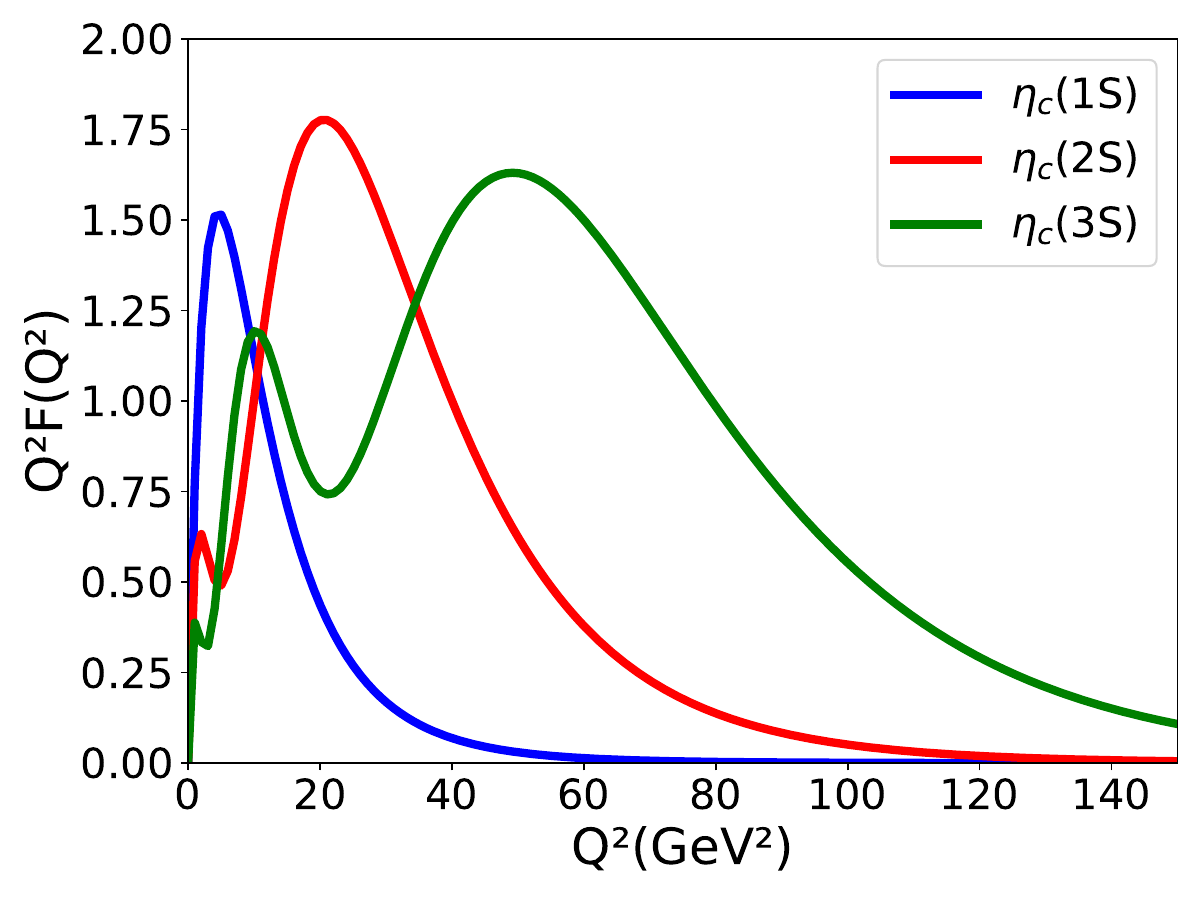}
    \subcaption{}
    \label{fig:Q2F_cc}
\end{subfigure}
\caption{EMFFs of charmonia. 
    (a) The form factors $F(Q^2)$ for the ground and excited states, showing the systematic increase in slope with higher radial excitations, whilst comparing the result from Lattice QCD B1 and C1\cite{Li:2020gau}. 
    (b) $Q^2 F(Q^2)$ for the same states, highlighting their behavior at large momentum transfer.}
\label{fig:charm_formfactor}
\end{figure}

Figure~\ref{fig:charm_formfactor} displays the EMFFs $F(Q^2)$ of charmonia and its radial excitations. The $\eta_c(1S)$ EMFF shows a steady decline in value as the momentum transfer $Q^2$ increases. 
We compare our results with the ground state $\eta_c(1S)$ values obtained by the B1 and C1 Lattice QCD \cite{Li:2020gau}. We see that at low momentum transfer, the Lattice QCD falls roughly in line with the ground state results with minor deviations. 
While the ground-state $\eta_c(1S)$ follows a gradual decline, the $\eta_c(2S)$ and $\eta_c(3S)$ states decrease more rapidly, indicating larger charge radii and spatial extensions, only to converge at a common value at around 10 GeV$^2$. 
The EMFFs for excited states at high momentum transfer $Q^2F(Q^2)$ exhibit oscillations due to the nodal structure of the wave functions.
The $\eta_c (1S)$ wave function has a maximum probability density near the origin with no internal zeros, forming no nodes. The $\eta_c (2S)$ wave function changes sign once, having an inner region which is positive and an outer region which is negative, forming one node. Finally, the $\eta_c (3S)$ state wave function changes sign twice, alternating between three regions of opposite signs, forming two nodes, lowering the overall EMFF.

The rms charge radii in Table~\ref{tab:charm_radii} 
display how the $c\bar{c}$ system grows as it experiences excitations. In our model, the rms charge radius increases steadily from 0.249(40) ~fm for the $\eta_c (1S)$ ground state, to 0.369(59) ~fm for the $\eta_c (2S)$ state and to 0.466(75) ~fm for the $\eta_c (3S)$ state. We see that there is a steady increase in each state, where the radius of $\eta_c (2S)$ grows roughly 1.48 times more than $\eta_c (1S)$, while the increase from $\eta_c (2S)$ to $\eta_c (3S)$ is around 1.26 times more. The total radius increase from the $\eta_c (1S)$ to $\eta_c (3S)$ states accumulates to about 1.87 times more. The results are as expected whereas the radial quantum number increases, the quark-antiquark pair spends more time at larger separations, and the meson becomes more spatially extended. The uncertainties found for all mesons take into account both parameter uncertainties and numerical effects on the EMFF. 
The larger errors in the $\eta_c (2S)$ and $\eta_c (3S)$ states come from their radial nodes, which amplify sensitivity near $Q^2 = 0$. However, the relative error across all three mesons for each state is similar, being around 16\% for charmonia, 20\% for $B_c$ mesons, and 13\% for bottomonia.

\begin{table}[t]
\tbl{Root mean square charge radii of the $\eta_c(1S)$, $\eta_c(2S)$, and $\eta_c(3S)$ states. The rms values are given in units of femtometers (fm).\label{tab:charm_radii}}
{\tabcolsep20pt\begin{tabular}{@{}llll@{}}
\toprule
$\sqrt{\expval{r^2}}$ & $\eta_c(1S)$ & $\eta_c(2S)$ & $\eta_c(3S)$ \\
\colrule
Our Model & 0.249(40) & 0.369(59) & 0.466(75) \\
Arifi et.al. \cite{Arifi:2022pal} & 0.205 & 0.359 & $\dots$ \\
Lattice QCD, B1 \cite{Li:2020gau} & 0.228(63) & $\dots$ & $\dots$ \\
Lattice QCD, C1 \cite{Li:2020gau} & 0.210(63) & $\dots$ & $\dots$ \\
Dudek et.al.(L.QCD) \cite{Dudek:2006ej} & 0.255 & $\dots$ & $\dots$ \\
Delaney et.al.(L.QCD) \cite{Delaney:2023fsc} & 0.243 & 0.620(179) & $\dots$ \\
Li et.al. \cite{Li:2015zda} & 0.195 & 0.384 & $\dots$ \\
Hern\'andez-Pinto et.al. \cite{Hernandez-Pinto:2023yin} & 0.200 & $\dots$ & $\dots$ \\
Adhikari et.al. \cite{Adhikari:2018umb} & 0.207(71) & 0.386(89) & $\dots$ \\
Serafin et.al. \cite{Serafin:2020egn} & 0.249 & 0.381 & $\dots$ \\
\botrule
\end{tabular}}
\end{table}

Compared to other studies, our results remain mostly consistent, as seen in Table~\ref{tab:charm_radii}. For the ground state $\eta_c (1S)$, our value is slightly higher than the results of Arifi \textit{et al.}~\cite{Arifi:2022pal}, but overall relatively similar. 
The results also vary slightly with other works such as BLFQ~\cite{Li:2015zda} and lattice QCD which supports the reliability of our result. 
For the $\eta_c (2S)$ state, our predicted value of 0.369~fm shows good agreement with the results of Arifi \textit{et al.}~\cite{Arifi:2022pal}, and is also consistent with BLFQ~\cite{Li:2015zda}(0.384~fm) and Adhikari \textit{et al.}~\cite{Adhikari:2018umb} (0.386(89)~fm). In contrast, the lattice QCD displays a larger value, around 0.620(179)~fm~\cite{Delaney:2023fsc}. This deviation can be correlated to the node present in the $\eta_c(2S)$ wave function, which suppresses the short-distance contribution and increases sensitivity to the long-range tail. The lattice QCD naturally produces a larger radius compared to other works, because the $\eta_c (2S)$ wave function led to high sensitivity to long range behavior in the extracted slope of the EMFF~\cite{Delaney:2023fsc}.
 Regarding the state $\eta_c (3S)$, our result serves as a prediction, since there are few data available for comparison. 
However, this state lies above the open charm threshold, where coupling to meson-meson channels can affect its EMFF.



\subsection{$B_c$ meson}

\begin{figure}[t!]
\centering
\begin{subfigure}{0.48\textwidth}
    \centering
    \includegraphics[width=\linewidth]{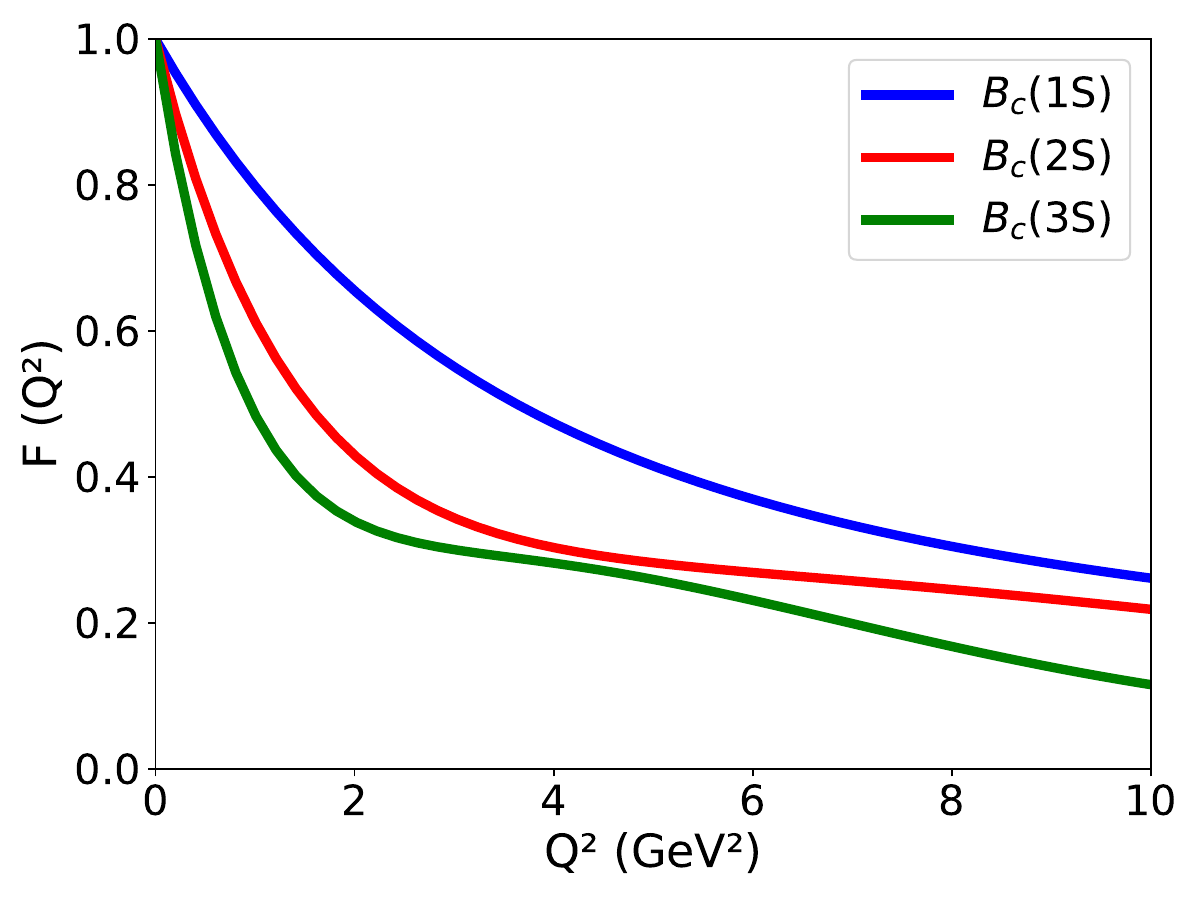}
    \subcaption{}
\end{subfigure}%
\hfill
\begin{subfigure}{0.48\textwidth}
    \centering
    \includegraphics[width=\linewidth]{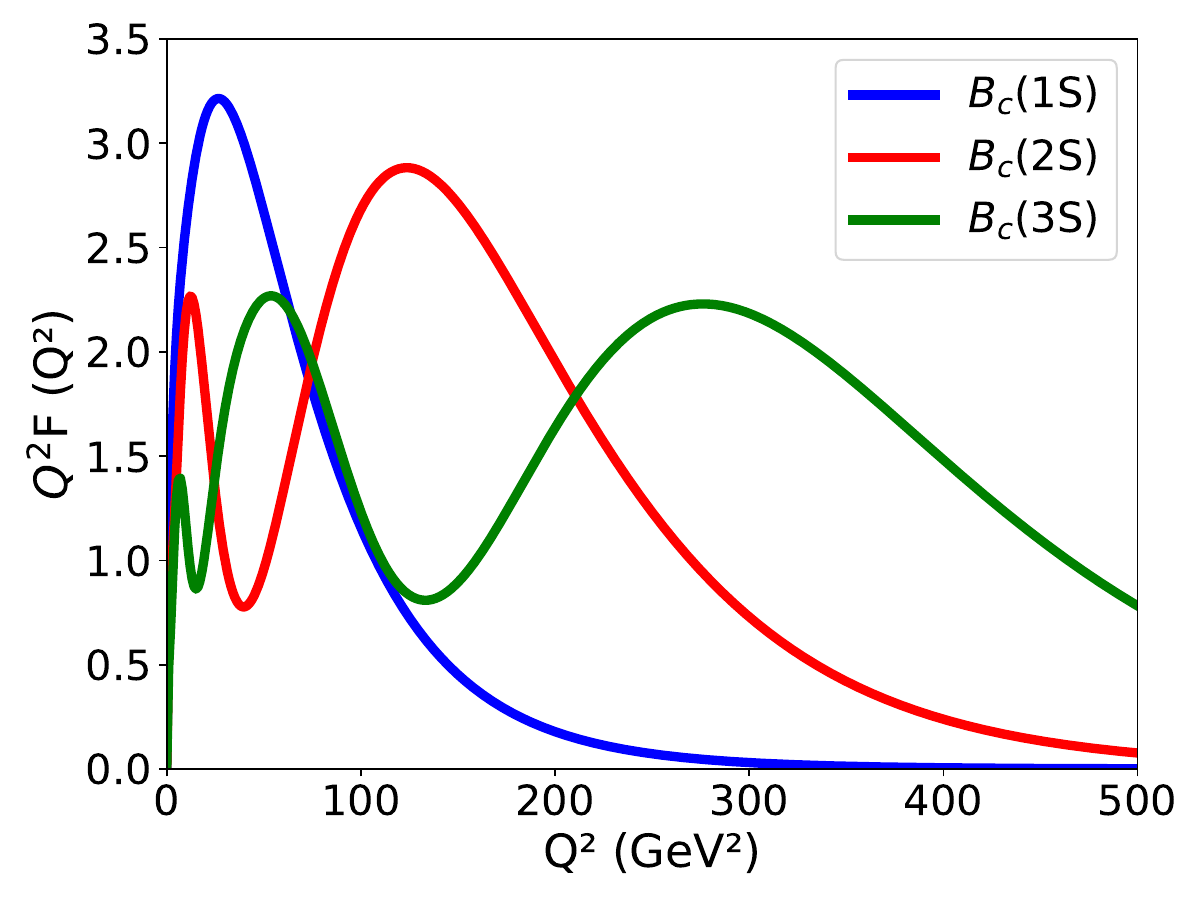}
    \subcaption{}
\end{subfigure}
\caption{EMFFs of the $B_c$ meson. 
    (a) The form factors $F(Q^2)$ for the ground and excited states, showing the systematic increase in slope with higher radial excitations. 
    (b) The form factors $Q^2 F(Q^2)$ for the same states, highlighting their behaviors at large momentum transfer.}
\label{fig:bc_formfactor}
\end{figure}

The EMFF results for the $B_c$ meson (Fig.~\ref{fig:bc_formfactor}) feature a similar trend to those of $\eta_c$, with each state of form factor gradually decreasing as $Q^2$ increases, though the $B_c$ form factor  fall more slowly, indicating a more compact internal structure. 
The $B_c (1S)$ EMFF stays in good agreement with a compact configuration, while the $B_c (2S)$ and $B_c (3S)$ states fall more rapidly, as they are larger and more diffuse, but show a different trend than those of charmonia and bottomonia due to its flavor asymmetry. The $B_c$ meson EMFFs still exhibit growth with radial excitation, showing that the excited states are dependent on the model, which highlights the need to capture the non-perturbative behavior of QCD. 
At higher momentum transfer, the EMFFs also show some oscillations due to the nodal structure of their wave functions. 
The $B_c (3S)$ wave functions, however, behaves differently than the 3S states of $\eta_c$ and $\eta_b$ because of their unequal quark masses. This imbalance causes the radial nodes to be asymmetrically distributed, leading to even stronger cancellations between the $\eta_c$ and $\eta_b$ quark contributions. 

The rms charge radii of the $B_c$ meson displays the interaction between the heavy bottom quark and the lighter charm quark. The rms increases steadily from 0.235(47)~fm for the $B_c (1S)$ ground state, 0.355(71)~fm for the $B_c (2S)$ state, and to 0.447(89)~fm for the $B_c (3S)$ state as shown in Table~\ref{tab:bc_radii}. These values are smaller than those of charmonia, reflecting the stronger binding and heavier reduced mass of the $b\bar{c}$ system. The radius grows by about a factor of 1.51 times from $B_c(1S)$ to $B_c(2S)$, by roughly 1.26 times from $B_c(2S)$ to $B_c(3S)$, and by about 1.9 times from $B_c(1S)$ to $B_c(3S)$. This steady growth with excitation reflects how higher radial excitations drive the quark and antiquark further apart, thereby increasing the effective size of the meson. 

In comparison with earlier studies, our results are generally similar to other works with minor differences, as seen in Table ~\ref{tab:bc_radii} 
For the $B_c (1S)$ state, our value of 0.235(47) ~fm is slightly larger than those of Arifi \textit{et al.}~\cite{Arifi:2022pal} and Hern\'andez-Pinto \textit{et al.}~\cite{Hernandez-Pinto:2023yin}, 
which lie in the range 0.170-0.207~fm but smaller than the result from Serafin \textit{et.al.}~\cite{Serafin:2020egn}, which is  0.281~fm. 
The relatively small radius of the $B_c(1S)$ state is expected, since the strong attraction between the heavy (b) and (c) quarks keeps the system tightly bound and compact in size. This compactness also highlights how the heavier bottom quark dominates the center of mass, while the lighter charm quark exhibits more motion, producing a radius that lies between the more compact bottomonia and the more diffuse charmonia systems.

For the $B_c (2S)$ state, we compare our result of 0.355(71)~fm which is slightly smaller than the values from Arifi \textit{et al.}~\cite{Arifi:2022pal} (0.358~fm) and Serafin \textit{et al.}~\cite{Serafin:2020egn} (0.430~fm). Our result falls between the values reported in other studies, reflecting a balanced treatment of short and long range dynamics within the model. The small discrepancies may originate from the sensitivity of the excited-state LFWFs and the presence of a radial node in the $B_c(2S)$ state, which enhances the long range contributions. For the $B_c(3S)$ state, our result serves as a prediction, as no comparative data are currently available. 

\begin{table}[t!]
\centering
\tbl{Root mean square charge radii of the $B_c(1S)$, $B_c(2S)$, and $B_c(3S)$ states. The rms values are given in units of femtometers (fm).\label{tab:bc_radii}}
{\tabcolsep20pt\begin{tabular}{@{}llll@{}}
\toprule
$\sqrt{\expval{r^2}}$ & $B_c(1S)$ & $B_c(2S)$ & $B_c(3S)$ \\
\colrule
Our model & 0.235(47) & 0.355(71) & 0.447(89) \\
Arifi et.al. \cite{Arifi:2022pal} & 0.190 & 0.358 & $\dots$ \\
Hern\'andez-Pinto et.al. \cite{Hernandez-Pinto:2023yin} & 0.170 & $\dots$ & $\dots$ \\
Serafin et.al. \cite{Serafin:2020egn} & 0.281 & 0.430 & $\dots$ \\
\botrule
\end{tabular}}
\end{table}


\subsection{Bottomonia}

\begin{figure}[t!]
\centering
\begin{subfigure}{0.48\textwidth}
    \centering
    \includegraphics[width=\linewidth]{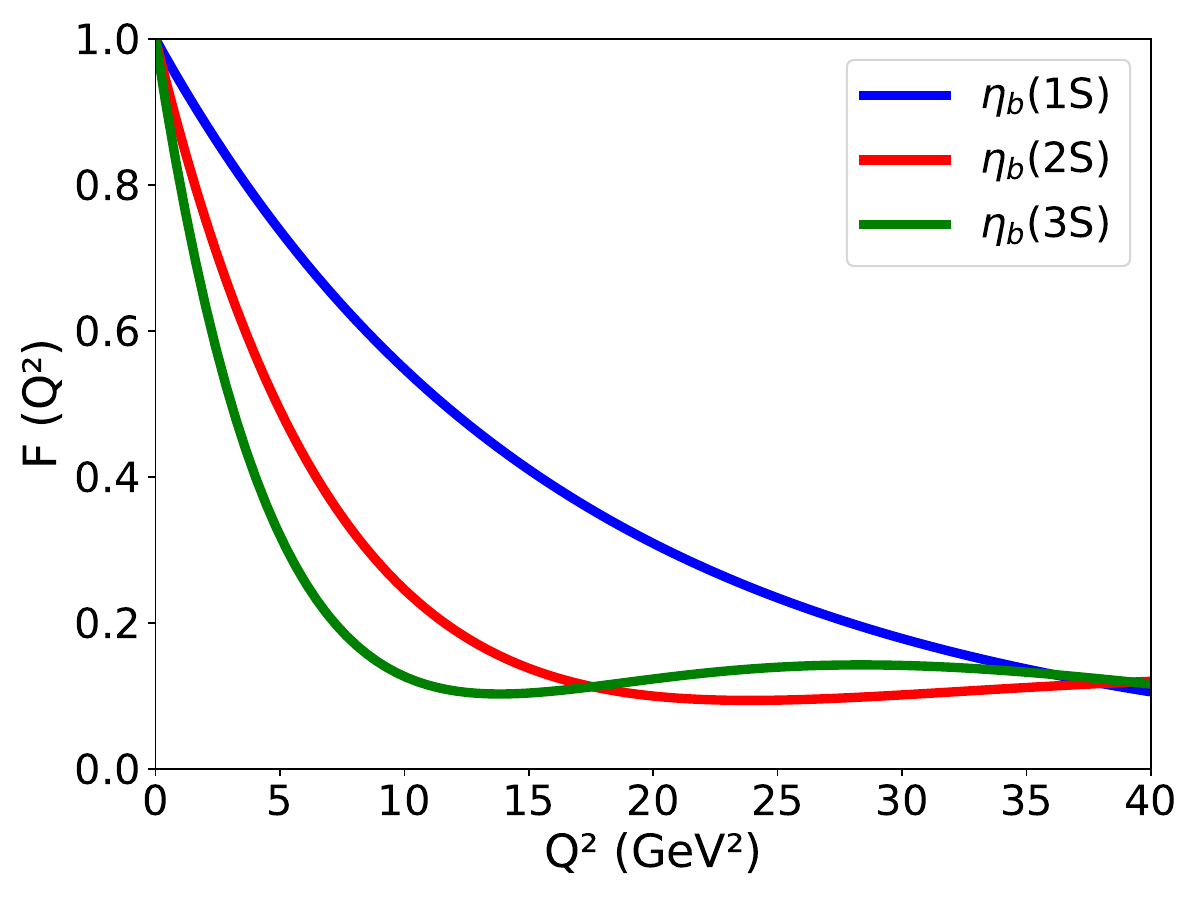}
    \subcaption{}
\end{subfigure}%
\hfill
\begin{subfigure}{0.48\textwidth}
    \centering
    \includegraphics[width=\linewidth]{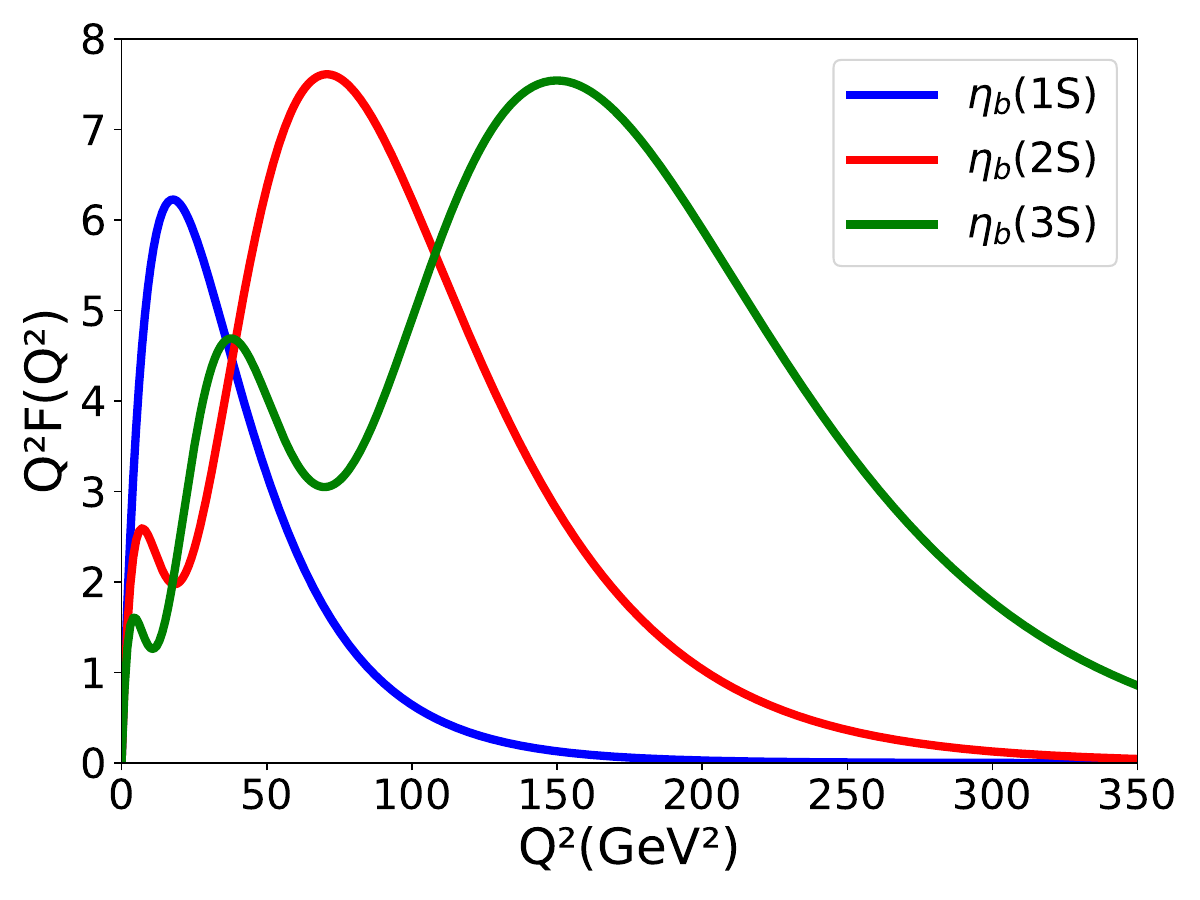}
    \subcaption{}
\end{subfigure}
\caption{EMFFs of bottomonia. (a) The form factors $F(Q^2)$ for the ground and excited states, showing the systematic increase in slope with higher radial excitations. (b) The form factors $Q^2 F(Q^2)$ for the same states, highlighting their behavior at large momentum transfer.}
\label{fig:bottom_formfactor}
\end{figure}

The EMFF results for bottomonia (Fig.~\ref{fig:bottom_formfactor}) follow a similar pattern to that of the charmonia system, but with a slower falloff at low $Q^2$, reflecting the higher mass and tighter binding of $b\bar{b}$. The EMFFs tend to converge like charmonia, the only difference being the convergence occurs at high momentum transfer around 36 GeV$^2$, while the charmonia converge at around 10 GeV$^2$. The $\eta_b(1S)$ EMFF decreases gradually, being more compact in its spatial configuration, while the $\eta_b(2S)$ and $\eta_b(3S)$ states tend to fall off more rapidly, indicating larger charge radii and more extended spatial distributions. 
The $\eta_b (1S)$ has no nodes, while the $\eta_b(2S)$ and $\eta_b(3S)$ contain one and two radial nodes, respectively, which introduce sign changes in their wave functions. 
These sign changes cause partial cancellations in the charge distribution, making the $\eta_b(2S)$ and $\eta_b(3S)$ form factors decrease more rapidly and exhibit mild oscillatory behavior compared to the $\eta_b(1S)$ state, though the overall suppression is less dramatic than in charmonia due to the tighter spatial confinement of bottomonia.

The rms charge radii of $\eta_b$ shown in Table~\ref{tab:bottom_radii}, displays
our model's results in radii of 0.118(16)~fm for the $\eta_b(1S)$ ground state, 0.179(24)~fm for the $\eta_b(2S)$ state, and 0.226(32)~fm for the $\eta_b(3S)$ state. These values indicate the expected steady increase with radial excitation, where higher states have LFWF that extend further out due to additional nodes. Compared to the $B_c$ and charmonia systems, bottomonia remain the most compact, reflecting the stronger binding between two heavy $b$ quarks.
There is an increase in the rms charge radius from $\eta_b(1S)$ to $\eta_b(2S)$, with the latter being about 1.51 times larger. This trend is in good agreement with the $B_c$ system from $B_c(1S)$ to $B_c(2S)$. The change from $\eta_b(2S)$ to $\eta_b(3S)$ also closely follows the $B_c$ pattern, with $\eta_b(3S)$ being about 1.26 times larger. Overall, the radius increases by about a factor of 1.91 from $\eta_b(1S)$ to $\eta_b(3S)$, supporting that higher radial excitations drive the quark and antiquark farther apart, enlarging the size of the meson. 

Our results are largely consistent with only minor differences with other models, as shown in Table~\ref{tab:bottom_radii} 
For the $\eta_b(1S)$ state, our value of 0.118(16)~fm is slightly higher than those of Arifi \textit{et al.}~\cite{Arifi:2022pal} (0.095~fm) and Hernández-Pinto \textit{et al.}~\cite{Hernandez-Pinto:2023yin} (0.071~fm), but close to Li \textit{et al.}~\cite{Li:2015zda} (0.121(28)~fm) and Adhikari \textit{et al.}~\cite{Adhikari:2018umb} (0.126(32)~fm). 
This small spread comes from relativistic effects and the short range potential for each model, as the compact $\eta_b(1S)$ state is more sensitive to these details.
For the $\eta_b(2S)$ state, our prediction of 0.179(24)~fm aligns with Arifi \textit{et al.}~\cite{Arifi:2022pal} (0.173~fm) and Serafin \textit{et al.}~\cite{Serafin:2020egn} (0.232~fm), and remains within the uncertainty of Li \textit{et al.}~\cite{Li:2015zda} (0.226(28)~fm) and Adhikari \textit{et al.}~\cite{Adhikari:2018umb} (0.237(45)~fm). The small variation between these results can be caused by the value of the bottom quark mass used in the calculation and the node in the $\eta_b(2S)$ LFWF, which increases sensitivity to the long range tail of the potential. 
For the $\eta_b(3S)$ state, our model gives a radius of 0.226(32)~fm, increasing size with excitation. 
Since there are currently no available results for this state, our value serves as a prediction for future theoretical and lattice studies. 

\begin{table}[t!]
\centering
\tbl{Root Mean square charge radii of the $\eta_b(1S)$, $\eta_b(2S)$, and $\eta_b(3S)$ states. The rms values are given in units of femtometers (fm). \label{tab:bottom_radii} } 
{\tabcolsep20pt\begin{tabular}{@{}llll@{}}
\toprule
$ \sqrt{\expval{r^2}}$ & $\eta_b(1S)$ & $\eta_b(2S)$ & $\eta_b(3S)$ \\
\colrule
Our model  & 0.118(16) & 0.179(24) & 0.226(32) \\
Arifi et.al. \cite{Arifi:2022pal} & 0.095 & 0.173 & $\dots$ \\
Li et.al, \cite{Li:2015zda} & 0.121(28) & 0.226(28) & $\dots$ \\
Hern\'andez-Pinto et.al. \cite{Hernandez-Pinto:2023yin} & 0.071 & $\dots$ & $\dots$ \\
Adhikari et.al. \cite{Adhikari:2018umb} & 0.126(32) & 0.237(45) & $\dots$ \\
Serafin et.al. \cite{Serafin:2020egn} & 0.152 & 0.232 & $\dots$ \\
\botrule
\end{tabular}}
\end{table}


\section{Conclusion}
\label{sec:conclusion}

In this work, we have investigated the EMFFs and charge radii of heavy quarkonia systems, namely charmonia, bottomonia, and the $B_c$ meson, within the framework of the light-front quark model. 
The model parameters were obtained from our previous study using the QCD-motivated quark model Hamiltonian~\cite{Ridwan:2024ngc}, and the light-front wave functions were matched to potential model solutions for each radial excitation. 

Our results show the expected pattern of spatial growth with increasing radial excitation: the $1S$ states are the most compact, while the $2S$ and $3S$ states become progressively larger due to the appearance of radial nodes in their wave functions. 
Among the systems studied, bottomonia are found to be the most tightly bound, charmonia being the most diffuse, 
and the $B_c$ meson sits between them, reflecting the interplay between its bottom and charm quark components. Additionally, our results are roughly consistent with experimental data and other theoretical models such as lattice QCD, BLFQ, and also BSE.

As for future studies, a more realistic description of LFWFs~\cite{Arifi:2024mff} is interesting to further refine our understanding of the meson form factors.
Some extensions for the form factor modifications in extreme environment such as in the magnetic field~\cite{Arifi:2025ivt}. 

\section*{Acknowledgments}

This work has been supported by the PUTI Q1 Grant from the University of Indonesia under Contract No. PKS-206/UN2.RST/HKP.05.00/2025. M.R. was supported by the PhD studentship jointly funded by the Science and Technology Facilities Council (STFC) under the UK Research and Innovation (UKRI) and the School of Engineering, Computing and Mathematics, University of Plymouth under work order No. GD105282-101. A.J.A. was supported by the JAEA Postdoctoral Fellowship Program and partly by the RCNP Collaboration Research Network Program under Project No. COREnet 057.

\section*{ORCID}
\noindent Rayn Rasyid Harjapradipta \orcid{0009-0000-2364-9601} \url{https://orcid.org/0009-0000-2364-9601}\\
\noindent Muhammad Ridwan \orcid{0000-0002-2949-5866} \url{https://orcid.org/0000-0002-2949-5866}\\
\noindent Ahmad Jafar Arifi \orcid{0000-0002-9530-8993} \url{https://orcid.org/0000-0002-9530-8993}\\
\noindent Terry Mart \orcid{0000-0003-4628-2245} \url{https://orcid.org/0000-0003-4628-2245}\\

\bibliographystyle{ws-mpla}
\bibliography{references}

@article{Gross:2022hyw,
    author = "Gross, Franz and others",
    title = "{50 Years of Quantum Chromodynamics}",
    doi = "10.1140/epjc/s10052-023-11949-2",
    journal = "Eur. Phys. J. C",
    volume = "83",
    pages = "1125",
    year = "2023"
}

@article{E598:1974sol,
    author = "Aubert, J. J. and others",
    collaboration = "E598",
    title = "{Experimental Observation of a Heavy Particle $J$}",
    reportNumber = "COO-3069-271",
    doi = "10.1103/PhysRevLett.33.1404",
    journal = "Phys. Rev. Lett.",
    volume = "33",
    pages = "1404--1406",
    year = "1974"
}

@article{Brambilla:2010cs,
    author = "Brambilla, N. and others",
    title = "{Heavy Quarkonium: Progress, Puzzles, and Opportunities}",
    reportNumber = "SLAC-R-996, TUM-EFT-11-10, CLNS-10-2066, ANL-HEP-PR-10-44, ALBERTA-THY-11-10, CP3-10-37, FZJ-IKP-TH-2010-24, INT-PUB-10-059, JLAB-THY-11-1308, FERMILAB-PUB-10-737-T",
    doi = "10.1140/epjc/s10052-010-1534-9",
    journal = "Eur. Phys. J. C",
    volume = "71",
    pages = "1534",
    year = "2011"
}

@article{Eichten:2007qx,
    author = "Eichten, Estia and Godfrey, Stephen and Mahlke, Hanna and Rosner, Jonathan L.",
    title = "{Quarkonia and their transitions}",
    reportNumber = "CLNS-07-1988, EFI-06-15, FERMILAB-PUB-07-006-T",
    doi = "10.1103/RevModPhys.80.1161",
    journal = "Rev. Mod. Phys.",
    volume = "80",
    pages = "1161--1193",
    year = "2008"
}

@article{Voloshin:2007dx,
    author = "Voloshin, M. B.",
    title = "{Charmonium}",
    reportNumber = "FTPI-MINN-07-34, UMN-TH-2625-07",
    doi = "10.1016/j.ppnp.2008.02.001",
    journal = "Prog. Part. Nucl. Phys.",
    volume = "61",
    pages = "455--511",
    year = "2008"
}

@article{Lewis:2012bh,
    author = "Lewis, Randy and Woloshyn, R. M.",
    title = "{More about excited bottomonium radiative decays}",
    doi = "10.1103/PhysRevD.86.057501",
    journal = "Phys. Rev. D",
    volume = "86",
    pages = "057501",
    year = "2012"
}

@article{Li:2020gau,
    author = "Li, Ning and Liu, Chao-Chen and Wu, Ya-Jie",
    title = "{Lattice study of form factors for charmonium}",
    doi = "10.1140/epja/s10050-020-00253-2",
    journal = "Eur. Phys. J. A",
    volume = "56",
    number = "9",
    pages = "242",
    year = "2020"
}

@article{Dudek:2006ej,
    author = "Dudek, Jozef J. and Edwards, Robert G. and Richards, David G.",
    title = "{Radiative transitions in charmonium from lattice QCD}",
    reportNumber = "JLAB-THY-06-457",
    doi = "10.1103/PhysRevD.73.074507",
    journal = "Phys. Rev. D",
    volume = "73",
    pages = "074507",
    year = "2006"
}

@article{Delaney:2023fsc,
    author = "Delaney, James and Thomas, Christopher E. and Ryan, Sin{\'e}ad M.",
    collaboration = "Hadron Spectrum",
    title = "{Radiative transitions in charmonium from lattice QCD}",
    doi = "10.1007/JHEP05(2024)230",
    journal = "JHEP",
    volume = "05",
    pages = "230",
    year = "2024"
}

@article{Choi:2009ai,
    author = "Choi, Ho-Meoyng and Ji, Chueng-Ryong",
    title = "{Semileptonic and radiative decays of the B(c) meson in light-front quark model}",
    doi = "10.1103/PhysRevD.80.054016",
    journal = "Phys. Rev. D",
    volume = "80",
    pages = "054016",
    year = "2009"
}

@article{Choi:2007se,
    author = "Choi, Ho-Meoyng",
    title = "{Decay constants and radiative decays of heavy mesons in light-front quark model}",
    doi = "10.1103/PhysRevD.75.073016",
    journal = "Phys. Rev. D",
    volume = "75",
    pages = "073016",
    year = "2007"
}

@article{JeffersonLab:2008jve,
    author = "Huber, G. M. and others",
    collaboration = "Jefferson Lab",
    title = "{Charged pion form-factor between Q**2 = 0.60-GeV**2 and 2.45-GeV**2. II. Determination of, and results for, the pion form-factor}",
    reportNumber = "JLAB-PHY-08-864",
    doi = "10.1103/PhysRevC.78.045203",
    journal = "Phys. Rev. C",
    volume = "78",
    pages = "045203",
    year = "2008"
}

@article{Amendolia:1984nz,
    author = "Amendolia, S. R. and others",
    title = "{A Measurement of the Pion Charge Radius}",
    reportNumber = "CERN-EP/84-59",
    doi = "10.1016/0370-2693(84)90655-5",
    journal = "Phys. Lett. B",
    volume = "146",
    pages = "116--120",
    year = "1984"
}

@article{Dirac:1949cp,
    author = "Dirac, Paul A. M.",
    title = "{Forms of Relativistic Dynamics}",
    doi = "10.1103/RevModPhys.21.392",
    journal = "Rev. Mod. Phys.",
    volume = "21",
    pages = "392--399",
    year = "1949"
}

@article{Brodsky:1997de,
    author = "Brodsky, Stanley J. and Pauli, Hans-Christian and Pinsky, Stephen S.",
    title = "{Quantum chromodynamics and other field theories on the light cone}",
    reportNumber = "SLAC-PUB-7484, MPIH-V1-1997",
    doi = "10.1016/S0370-1573(97)00089-6",
    journal = "Phys. Rept.",
    volume = "301",
    pages = "299--486",
    year = "1998"
}

@article{Arifi:2022pal,
    author = "Arifi, Ahmad Jafar and Choi, Ho-Meoyng and ji, Chueng-Ryong and Oh, Yongseok",
    title = "{Mixing effects on 1S and 2S state heavy mesons in the light-front quark model}",
    doi = "10.1103/PhysRevD.106.014009",
    journal = "Phys. Rev. D",
    volume = "106",
    number = "1",
    pages = "014009",
    year = "2022"
}

@article{Choi:2024ptc,
	author       = "Choi, Ho-Meoyng and Ji, Chueng-Ryong",
	title        = "{Consistency of the pion form factor and unpolarized transverse momentum dependent parton distributions beyond leading twist in the light-front quark model}",
	doi          = "10.1103/PhysRevD.110.014006",
	journal      = "Phys. Rev. D",
	volume       = 110,
	number       = 1,
	pages        = "014006",
	year         = 2024,
}

@article{Li:2015zda,
    author = "Li, Yang and Maris, Pieter and Zhao, Xingbo and Vary, James P.",
    title = "{Heavy Quarkonium in a Holographic Basis}",
    doi = "10.1016/j.physletb.2016.04.065",
    journal = "Phys. Lett. B",
    volume = "758",
    pages = "118--124",
    year = "2016"
}

@article{Hernandez-Pinto:2023yin,
    author = "Hern{\'a}ndez-Pinto, R. J. and Guti{\'e}rrez-Guerrero, L. X. and Bashir, A. and Bedolla, M. A. and Higuera-Angulo, I. M.",
    title = "{Electromagnetic form factors and charge radii of pseudoscalar and scalar mesons: A comprehensive contact interaction analysis}",
    reportNumber = "JLAB-THY-23-3748",
    doi = "10.1103/PhysRevD.107.054002",
    journal = "Phys. Rev. D",
    volume = "107",
    number = "5",
    pages = "054002",
    year = "2023"
}

@article{Adhikari:2018umb,
    author = "Adhikari, Lekha and Li, Yang and Li, Meijian and Vary, James P.",
    title = "{Form factors and generalized parton distributions of heavy quarkonia in basis light front quantization}",
    doi = "10.1103/PhysRevC.99.035208",
    journal = "Phys. Rev. C",
    volume = "99",
    number = "3",
    pages = "035208",
    year = "2019"
}

@article{Serafin:2020egn,
    author = "Serafin, Kamil",
    title = "{Form factors and structure functions of heavy mesons and baryons}",
    doi = "10.22323/1.374.0079",
    journal = "PoS",
    volume = "LC2019",
    pages = "079",
    year = "2020"
}

@article{Hwang:2001th,
    author = "Hwang, Chien-Wen",
    title = "{Charge radii of light and heavy mesons}",
    doi = "10.1007/s100520200904",
    journal = "Eur. Phys. J. C",
    volume = "23",
    pages = "585--590",
    year = "2002"
}

@article{Zhang:2024nxl,
    author = "Zhang, Jin-Li",
    title = "{{\ensuremath{\rho}} meson form factors and parton distribution functions in impact parameter space}",
    doi = "10.1088/1674-1137/adab61",
    journal = "Chin. Phys. C",
    volume = "49",
    number = "4",
    pages = "043104",
    year = "2025"
}

@article{Maris:2000sk,
    author = "Maris, Pieter and Tandy, Peter C.",
    title = "{The pi, K+, and K0 electromagnetic form-factors}",
    reportNumber = "KSU-CNR-106-00",
    doi = "10.1103/PhysRevC.62.055204",
    journal = "Phys. Rev. C",
    volume = "62",
    pages = "055204",
    year = "2000"
}

@article{Wang:2022mrh,
    author = "Wang, Xiaobin and Xing, Zanbin and Kang, Jiayin and Raya, Kh{\'e}pani and Chang, Lei",
    title = "{Pion scalar, vector, and tensor form factors from a contact interaction}",
    doi = "10.1103/PhysRevD.106.054016",
    journal = "Phys. Rev. D",
    volume = "106",
    number = "5",
    pages = "054016",
    year = "2022"
}

@article{Ridwan:2024ngc,
    author = "Ridwan, Muhammad and Arifi, Ahmad Jafar and Mart, Terry",
    title = "{Self-consistent M1 radiative transitions of excited Bc and heavy quarkonia with different polarizations in the light-front quark model}",
    doi = "10.1103/PhysRevD.111.016011",
    journal = "Phys. Rev. D",
    volume = "111",
    number = "1",
    pages = "016011",
    year = "2025"
}

@article{Choi:1997iq,
    author = "Choi, Ho-Meoyng and Ji, Chueng-Ryong",
    title = "{Mixing angles and electromagnetic properties of ground state pseudoscalar and vector meson nonets in the light cone quark model}",
    reportNumber = "NCSU-97-32",
    doi = "10.1103/PhysRevD.59.074015",
    journal = "Phys. Rev. D",
    volume = "59",
    pages = "074015",
    year = "1999"
}

@article{Choi:2015ywa,
    author = "Choi, Ho-Meoyng and Ji, Chueng-Ryong and Li, Ziyue and Ryu, Hui-Young",
    title = "{Variational analysis of mass spectra and decay constants for ground state pseudoscalar and vector mesons in the light-front quark model}",
    doi = "10.1103/PhysRevC.92.055203",
    journal = "Phys. Rev. C",
    volume = "92",
    number = "5",
    pages = "055203",
    year = "2015"
}

@article{Acharyya:2024tql,
    author = "Acharyya, Ritwik and Puhan, Satyajit and Dahiya, Harleen and Kumar, Narinder",
    title = "{Spectroscopy of excited quarkonium states in the light-front quark model}",
    doi = "10.1088/1674-1137/ad8ec3",
    journal = "Chin. Phys. C",
    volume = "49",
    number = "2",
    pages = "023104",
    year = "2025"
}

@article{Melosh:1974cu,
    author = "Melosh, H. J.",
    title = "{Quarks: Currents and constituents}",
    doi = "10.1103/PhysRevD.9.1095",
    journal = "Phys. Rev. D",
    volume = "9",
    pages = "1095",
    year = "1974"
}

@article{Arifi:2024mff,
    author = "Arifi, Ahmad Jafar and Happ, Lucas and Ohno, Shuhei and Oka, Makoto",
    title = "{Structure of heavy mesons in the light-front quark model}",
    doi = "10.1103/PhysRevD.110.014020",
    journal = "Phys. Rev. D",
    volume = "110",
    number = "1",
    pages = "014020",
    year = "2024"
}

@article{Arifi:2022qnd,
    author = "Arifi, Ahmad Jafar and Choi, Ho-Meoyng and Ji, Chueng-Ryong and Oh, Yongseok",
    title = "{Independence of current components, polarization vectors, and reference frames in the light-front quark model analysis of meson decay constants}",
    doi = "10.1103/PhysRevD.107.053003",
    journal = "Phys. Rev. D",
    volume = "107",
    number = "5",
    pages = "053003",
    year = "2023"
}

@article{Arifi:2025ivt,
    author = "Arifi, Ahmad Jafar and Suzuki, Kei",
    title = "{Structure of heavy quarkonia in a strong magnetic field}",
    doi = "10.1103/pwsl-xrq5",
    journal = "Phys. Rev. D",
    volume = "112",
    number = "9",
    pages = "094013",
    year = "2025"
}

\end{document}